\RequirePackage{fix-cm}
\documentclass{svjour2}                     
\smartqed  
\usepackage{graphicx}
\usepackage{amsmath}

 \journalname{my journal}
\begin{document}

\title{Exit probability in generalised kinetic Ising model
}


\author{Parna Roy         \and
        Parongama Sen 
}

\authorrunning{P. Roy \etal} 

\institute{P .Roy \at
             Department of Physics, University of Calcutta-
92 Acharya Prafulla Chandra Road, Kolkata 700009, India  \\
              \email{parna.roy14@gmail.com}           
           \and
           P. Sen \at
              Department of Physics, University of Calcutta-
92 Acharya Prafulla Chandra Road, Kolkata 700009, India\\
              \email{parongama@gmail.com}
}

\date{Received: date / Accepted: date}

\maketitle

\begin{abstract}
In this paper we study generalised Ising Glauber models with inflow of information in one dimension 
 and derive expressions for the exit probability using well established analytical methods. 
 The analytical expressions agree very well with 
 the results obtained from numerical simulation only when the interaction is restricted to the nearest neighbor.
 But as the range of interaction is increased the analytical results deviate from simulation results systematically. 
 The reasons for the deviation as well as some related open questions are discussed. 
\end{abstract}

\section{Introduction}
\label{intro}
Various physical systems and phenomena can be characterised using binary variables 
having values $\sigma=\pm1$ (or $0, 1$). The classic example is the Ising model
which has only two allowed spin states, taken as $\sigma=\pm1$. 
Ising spins can be used to represent the states 
of individual components of many other systems, particularly in models related to 
social phenomena \cite{liggett,socio,sociobook,galam1}. The voter model \cite{liggett} and Sznajd model \cite{sznajd}
for opinion formation in a society are typical examples.

When the dynamics of ordering is studied in such systems, one usually starts with a fraction $x$ of the variables 
having value $1$ and the rest equal to $-1$. $x=\frac{1}{2}$ corresponds to a completely random initial state. In one dimension,
usually all models with binary variables end up in a consensus state with all particle states having values $1$ (or $-1$). 
The two absorbing states occur with different probability. As a function of $x$, the probability that all 
particles end up in a state $1$ is called the exit probability $E(x)$.

 Recently a lot of effort has been put to identify the characteristic features 
 of exit probability for models in one dimension. While for the Ising model with Glauber dynamics,
 $E(x)=x$ is an exact result,
  in the Sznajd model the variation of $E(x)$ is nonlinear \cite{cast,slan,lamb}.
  Extending only to the 
  next nearest neighbor, the generalised Ising model studied in \cite{cast} showed that the results 
  were identical with the Sznajd model. This led to the idea that exit probability can have nonlinear 
  variations for both inflow and outflow dynamics (examples are the Ising and the Sznajd models respectively). Introducing further neighbor 
  interaction, asymmetry and fluctuations, the exit probability has been studied in Ising Glauber models \cite{cast,psp}; 
  once again nonlinear variations are observed. There is one school of thought that $E(x)$ has a step function 
  like behaviour for all one dimensional models \cite{galam}. However this has been observed only in certain 
  cases e.g. when one introduces a neighbouring domain size 
  dependent dynamics \cite{ssp,psp1} or in a mean field like approximation for the nonlinear $q$ voter model \cite{prado}.
  These recent results heve generated a lot of interest 
  in the study of exit probability in dynamical spin models.

  Most of the available results are obtained by numerical simulation when the range of interaction exceeds beyond nearest neighbor 
  and/or other features are introduced in the Ising Glauber (IG) model. Universal forms for $E(x)$ for 
  Ising Glauber models \cite{psp} and the nonlinear $q$ voter model \cite{pkm} have been 
  proposed although there is an existing controversy regarding its validity \cite{prado,new} for the latter. 
  For the so called generalised Ising Glauber models, 
  we have obtained an expression of $E(x)$ using Kirkwood approximation (KA) following \cite{kirk}. However, 
  an additional assumption has to be used along with KA to obtain the results \cite{slan,kirk}. Kirkwood approximation has proven 
  quite successful in a variety of application to reaction kinetics \cite{reac}. Also for Sznajd model this 
  approach 
  gives very good agreement with computer simulation results \cite{slan,lamb}.

To derive $E(x)$ using Kirkwood approximation (KA) one has to consider the master equation for the 
probability distribution of a given spin configuration $P(\sigma_1,\ldots\sigma_{N},t)$ as,
\begin{eqnarray*}
 \frac{d}{dt}P((\sigma_1,\ldots,\sigma_{N});t)&=&-\sum_jw(\sigma_j)P(\sigma_1,\ldots,\sigma_j,\ldots,\sigma_{N};t)\\
    &+&\sum_jw(-\sigma_j)P(\sigma_1,\ldots,-\sigma_j,\ldots,\sigma_{N};t),
\end{eqnarray*}
  where $w(\sigma_j)$ be the probability per unit time that the $j$th spin flips from 
$\sigma_j$ to $-\sigma_j$ which we refer to as flipping rate. 

Using this master equation the mean value of the $j$th spin
$$\langle\sigma_j\rangle=\sum_{\{\sigma\}}\sigma_jP({\sigma};t)$$ evolves as 
\begin{equation}
\label{rate1}
\frac{d}{dt}\langle\sigma_j\rangle=-2\langle w(\sigma_j)\sigma_j\rangle
\end{equation}
and the nearest-neighbor correlation function $\langle\sigma_j\sigma_{j+1}\rangle$ evolves as,
\begin{equation}
\label{rate2}
\frac{d}{dt}\langle\sigma_j\sigma_{j+1}\rangle =-2\langle\sigma_j\sigma_{j+1}\{w(\sigma_j)+w(\sigma_{j+1})\}\rangle.
\end{equation}
Because of spatial homogeneity all $\langle\sigma_j\rangle$ are identical and we write 
the magnetisation as $m\equiv\langle\sigma_j\rangle$. The two-spin correlation $\langle\sigma_j\sigma_{j+1}\rangle$ 
is written as $m_2$.

In the Kirkwood approximation the $3$-spin correlation function is decoupled as 
\begin{equation}
\label{ka1}
 \langle\sigma_{j-1}\sigma_j\sigma_{j+1}\rangle\approx mm_2
\end{equation}
   and the $4$-point 
  function is factorized as the product of $2$-
  point functions, 
  \begin{equation}
  \label{ka2}
 \langle\sigma_{j-1}\sigma_j\sigma_{j+1}\sigma_{j+2}\rangle\approx m_2^2.  
  \end{equation}
 We have used another approximation given in \cite{slan,kirk}. It is assumed that
  $\langle\sigma_i\sigma_{i+n}\rangle$ only weakly depends on distance $n$. This is justified 
 if the domains are large i.e. at later stages of the evolution.\\
 In this paper we needed to assume 
 \begin{equation}
 \label{ka3}
  \langle\sigma_i\sigma_{i+n}\rangle\equiv m_2
 \end{equation}
 for $1 \le n \le4$. The boundary conditions  at time $t=0$ and $t=\infty$ are 
$$m(0)=2x-1$$ and $$m(\infty)=2E(x)-1.$$
Since initially all the
spins are uncorrelated, we also have $m_2(0)=m(0)^2$. 

Here we have studied $E(x)$ for 
models with nearest and next nearest neighbor interaction in one dimension using Kirkwood approximation 
and the approximation given in equation (\ref{ka3}).
For all the models considered here, numerical results for $E(x)$ were reported in an 
earlier study \cite{psp} showing strong dependence on the model parameters. It is possible
to obtain approximate results for $E(x)$ in all these models using equations (\ref{ka1}-\ref{ka3}) 
and also compare them with the numerical ones.

   For each of the models considered, we evaluate 
$w(\sigma_i)$ at zero temperature. Solving the equations of motion and applying the above boundary 
conditions we obtain an expression of $E(x)$. In section II we calculate the exit probability for a generalised
IG model 
with nearest neighbor interaction and in section III we calculate the exit probability for generalised 
IG models with next nearest neighbor interaction. In section IV
we have made comparison 
with the simulation results. In section V the results are discussed and we
conclude that this method works very well for models with 
nearest neighbor interaction but not so good for models with long ranged interaction.

\section{$W_0$ model with nearest neighbor interaction}
\label{sec:1}
The generalised IG model with nearest neighbor interaction considered here is called the $W_0$ model \cite{godreche}.
    In the $W_0$ model at zero temperature, the central spin $\sigma_i$ is flipped 
with probability $W_0$ when the sum of the neighboring spins is zero, i.e. $\sigma_{i-1}+\sigma_{i+1}=0$. Except this 
fact it is exactly like IG model with nearest neighbor interaction. $W_0=1/2$ corresponds to 
original Glauber dynamics and $W_0=1$ 
is the Metropolis rule. $W_0=0$ is not allowed since this is the case of constrained zero temperature Glauber dynamics.

Here the flipping rate of  $\sigma_i$ can be written as,
\begin{equation}
\label{flipw}
 w(\sigma_i)=\frac{1}{4}[1+2W_0-\sigma_i(\sigma_{i+1}+\sigma_{i-1})+(1-2W_0)\sigma_{i+1}\sigma_{i-1}].
\end{equation}
The details of the derivation of the spin flip rate is given in Appendix \ref{Appendix A}.

Using the expression for transition rate in equation (\ref{rate1}) we have,
\begin{eqnarray}
 \frac{d}{dt}\langle\sigma_j\rangle=&-&\frac{1}{2}[\langle\sigma_j\rangle+2W_0\langle\sigma_j\rangle-\langle\sigma_{j+1}\rangle\nonumber\\
 &-&\langle\sigma_{j-1}\rangle+\langle\sigma_j\sigma_{j+1}\sigma_{j-1}\rangle-2W_0\langle\sigma_j\sigma_{j+1}\sigma_{j-1}\rangle].
\end{eqnarray}
From equation (\ref{rate2}) we have,
\begin{eqnarray}
\label{near2}
\frac{d}{dt}\langle\sigma_j\sigma_{j+1}\rangle =&-&\frac{1}{2}[2(1+2W_0)\langle\sigma_j\sigma_{j+1}\rangle-2\nonumber\\
&-&2\langle\sigma_{j+1}\sigma_{j-1}\rangle+2(1-2W_0)\langle\sigma_j\sigma_{j-1}\rangle]
\end{eqnarray}

 Using all the approximations mentioned in section I we have,

\begin{equation}
 \frac{dm}{dt}=\frac{1}{2}m(1-2W_0)(1-m_2)
 \label{rate3}
\end{equation}
and
\begin{equation}
 \frac{dm_2}{dt}=1-m_2.
 \label{rate4}
\end{equation}

First we solve for eq. (\ref{rate4}). This gives 
$$m_2=1-Ce^{-t},$$ where $C=1-m_2(0)$.  Now inserting this into eq. (\ref{rate3}) we have
for the average magnetisation,
\begin{equation}
 m=m(0)\exp\left(\frac{C}{2}(1-2W_0)\right)\exp\left(\frac{C}{2}(2W_0-1)e^{-t}\right).
\end{equation}

Using the expressions for $m(\infty)$ and $m(0)$ and after some straightforward algebra we find the exit probability as,
\begin{equation}
E(x)=\frac{1}{2}[1+(2x-1)e^{2x(1-x)(1-2W_0)}] .
\label{W_0}
\end{equation}

\section{Generalised Ising Glauber models with next nearest neighbor interaction}
\label{sec:2}
We consider two types of generalised IG model with next nearest neighbor interaction. The first type can 
be described by a Hamiltonian defined as,
\begin{equation}
 H=-J_1\sum_{i}\sigma_i\sigma_{i+1}-J_2\sum_i\sigma_i\sigma_{i+2}
 \label{hamiltonian}
\end{equation}
where $J_1$ is the interaction strength with nearest neighbor and $J_2$ is the interaction strength with next nearest neighbor 
and $J_1, J_2 >0$. This can be called ferromagnetic asymmetric next nearest model or FA model as in \cite{psp}.
 The dynamics are different for the three cases, $J_1=J_2$, $J_1>J_2$ and $J_1<J_2$. The cases $J_1=J_2$ 
 and $J_1>J_2$ correspond to $G(2)$ and $C_2$ models of reference \cite{psp}.
Under zero temperature dynamics here one can again calculate the flipping rate using Kirkwood approximation. We have studied the 
three different cases separately.

The second type of model is defined by a dynamical rule; actually we consider the $W_0$ model
 with next nearest neighbor interaction. Dynamical rules here are exactly like the case 
$J_1=J_2$ (\ref{hamiltonian}) except for the case $\sum_{j=1}^2[\sigma_{i+j}+\sigma_{i-j}]=0$. In this case
the central spin $\sigma_i$ is flipped with probability $W_0$.

\subsection{\bf Case I : FA model with $J_1=J_2$}
\label{sec:3}

The flipping rate of a spin at site $i$ for this case 
at zero temperature can be written as,

\begin{eqnarray}
\label{flipfa}
w(\sigma_i)=&\frac{1}{2}&[1-\frac{3}{8}\sigma_i(\sigma_{i-2}+\sigma_{i-1}+\sigma_{i+2}+\sigma_{i+1})\nonumber\\
&+&\frac{1}{8}\sigma_i(\sigma_{i+1}\sigma_{i+2}\sigma_{i-2}+\sigma_{i-1}\sigma_{i+2}\sigma_{i-2}\nonumber\\
&+&\sigma_{i+1}\sigma_{i-1}\sigma_{i-2}+\sigma_{i+1}\sigma_{i+2}\sigma_{i-1})].
\end{eqnarray}
The detailed derivation of equation (\ref{flipfa}) is given in Appendix \ref{Appendix B}.

Using the above expression for transition rate in equation (\ref{rate1}) we have,
\begin{eqnarray}
 \frac{d}{dt}\langle\sigma_j\rangle=&-&[\langle\sigma_j\rangle-\frac{3}{8}(\langle\sigma_{j-2}\rangle+\langle\sigma_{j-1}\rangle+\langle\sigma_{j+1}\rangle+\langle\sigma_{j+2}\rangle)\nonumber\\
 &+&\frac{1}{8}(\langle\sigma_{j+1}\sigma_{j+2}\sigma_{j-2}\rangle+\langle\sigma_{j+2}\sigma_{j-2}\sigma_{j-1}\rangle\nonumber\\
 &+&\langle\sigma_{j-2}\sigma_{j-1}\sigma_{j+1}\rangle+\langle\sigma_{j-1}\sigma_{j+1}\sigma_{j+2}\rangle)].
\end{eqnarray}

From equation (\ref{rate2}) we have,
\begin{eqnarray}
\label{nnn}
\frac{d}{dt}\langle\sigma_j\sigma_{j+1}\rangle =&-&[2\langle\sigma_j\sigma_{j+1}\rangle-\frac{3}{4}(\langle\sigma_{j+1}\sigma_{j-2}\rangle+\langle\sigma_{j+1}\sigma_{j-1}\rangle+1\nonumber\\
&+&\langle\sigma_{j+1}\sigma_{j+2}\rangle)+\frac{1}{4}(\langle\sigma_{j+2}\sigma_{j-2}\rangle+\langle\sigma_{j+2}\sigma_{j-2}\sigma_{i+1}\sigma_{i-1}\rangle\nonumber\\
&+&\langle\sigma_{j-2}\sigma_{j-1}\rangle+\langle\sigma_{j-1}\sigma_{j+2}\rangle)].
\end{eqnarray}

Proceeding as before we have the rate equations for $m$ and $m_2$ as,
 \begin{equation}
  \frac{dm}{dt}=\frac{m}{2}(1-m_2)
  \label{grate3}
 \end{equation}

 and
 
 \begin{equation}
  \frac{dm_2}{dt}=-\frac{1}{4}[(1+m_2)^2-4].
  \label{grate4}
 \end{equation}

Solving eq. (\ref{grate4}) and using the initial conditions we have,
$$m_2=\frac{3e^{-t}+C}{C-e^{-t}},$$ where $C=\frac{m_2(0)+3}{m_2(0)-1}$. Now inserting this into eq. (\ref{grate3}) we get
for the average magnetisation,
\begin{equation}
 m=\frac{16m(0)}{(m(0)^2-1)^2}\frac{1}{(C-e^{-t})^2}.
\end{equation}

The exit probability is obtained following some simple algebraic steps as,
\begin{equation}
 E(x)=\frac{x^4-2x^3+3x^2}{2(x^2-x+1)^2}.
\label{exit}
\end{equation}

\begin{figure}
\hspace{1.0cm}
\includegraphics[width=9.5cm,angle=0]{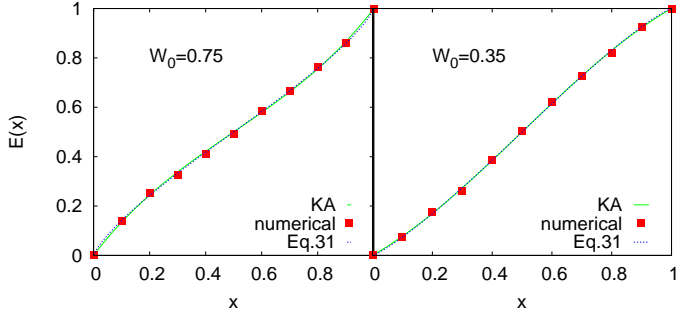}
\caption{Exit probability for $W_0$ model with nearest neighbor interaction with $W_0=0.75$ (left)
and $W_0=0.35$ (right). KA indicates Kirkwood approximating in all the diagrams.}
\label{wmodel}
\end{figure}

\begin{figure}
\hspace{1.0cm}
 \includegraphics[width=9.5cm,angle=0]{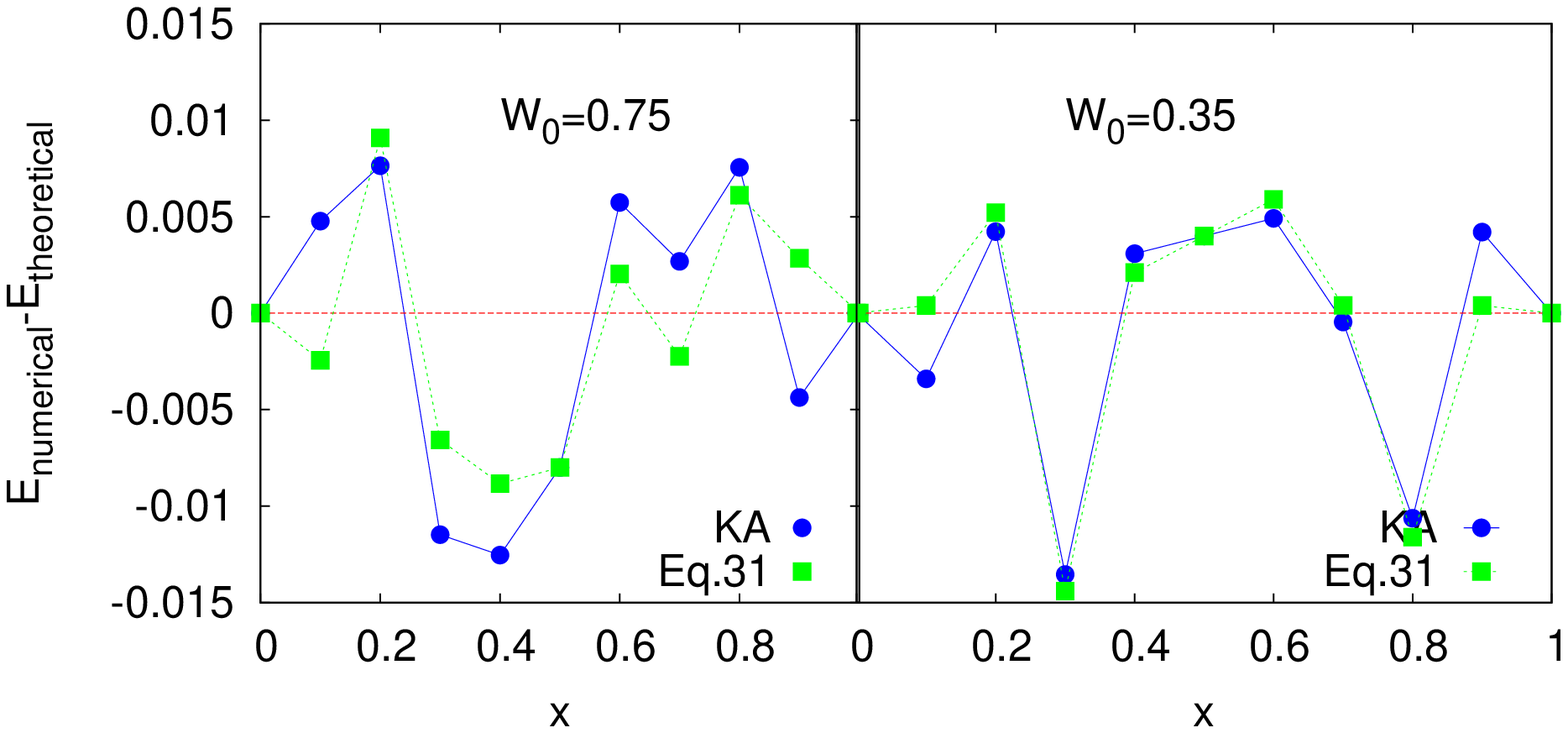}
\caption{Difference between the exit probability obtained from numerical simulation and
the one predicted by eq. (\ref{W_0}) and eq. (\ref{alpha}) for the $W_0$ model with nearest neighbor interaction.}
\label{wmodel1}
\end{figure}   

\subsection{\bf Case II : FA model with $J_1>J_2$}  
\label{sec:4}
In this case the flipping rate for the $i$th spin can be written as,
\begin{eqnarray*}
 w(\sigma_i)=&\frac{1}{2}&[1-\frac{1}{2}\sigma_i(\sigma_{i+1}+\sigma_{i-1})-\frac{1}{4}\sigma_i(\sigma_{i+2}+\sigma_{i-2})\\
 &+&\frac{1}{4}\sigma_i\sigma_{i+1}\sigma_{i-1}(\sigma_{i+2}+\sigma_{i-2})].
\end{eqnarray*}

Using this transition rate and following the same procedure  the  
rate equations for the average magnetisation $m$ and nearest-neighbor spin correlation $m_2$ are obtained as,
\begin{equation}
 \frac{dm}{dt}=\frac{1}{2}m(1-m_2)
 \label{fa2}
\end{equation}
and
\begin{equation}
 \frac{dm_2}{dt}=1-m_2 .
 \label{fa1}
\end{equation}

Solving eq. (\ref{fa1}) we have $m_2=1-Ce^{-t}$, where $C=1-m_2(0)$. Hence we get from eq. (\ref{fa2}) 
$$m=m(0)\exp\left(\frac{1-m(0)^2}{2}\right)\frac{1}{\exp(\frac{C}{2}e^{-t})}.$$

After some straightforward steps we have,
\begin{equation}
 E(x)=\frac{1}{2}\left[1+(2x-1)e^{2x(1-x)}\right].
\label{exit1}
\end{equation}

\subsection{\bf Case III : FA model with $J_1<J_2$}
\label{sec:5}
  For FA model with $J_1<J_2$, the flipping rate for the $i$th spin at zero temperature is given by,
 
\begin{eqnarray*}
 w(\sigma_i)=&\frac{1}{2}&[1-\frac{1}{4}\sigma_i(\sigma_{i+1}+\sigma_{i-1})-\frac{1}{2}\sigma_i(\sigma_{i+2}+\sigma_{i-2})\\
 &+&\frac{1}{4}\sigma_i\sigma_{i+2}\sigma_{i-2}(\sigma_{i+1}+\sigma_{i-1})].
\end{eqnarray*}

The rate equations for this case are as follows,
\begin{equation}
 \frac{dm}{dt}=\frac{1}{2}m(1-m_2)
 \label{fa3}
\end{equation}
and
\begin{equation}
 \frac{dm_2}{dt}=\frac{1}{2}(1-m_2^2).
 \label{fa4}
\end{equation}

Solving eq. (\ref{fa4}) we have $m_2=\frac{e^t-C}{e^t+C}$, where $C=\frac{1-m_2(0)}{1+m_2(0)}$. 
Inserting $m_2$ in eq. (\ref{fa3}) we have 
$$m=m(0)\frac{1+C}{1+Ce^{-t}},$$

and we finally get 

\begin{equation}
 E(x)=\frac{x^2}{x^2+(1-x)^2}.
\label{exit2}
\end{equation}

\begin{figure}
\hspace{1.0cm}
\includegraphics[width=9.5cm,angle=0]{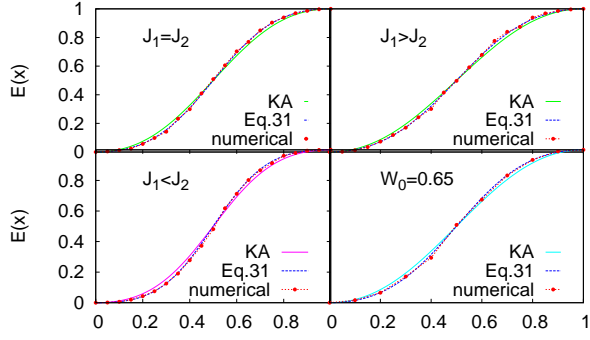}
\caption{Exit probability for models with next nearest neighbor interaction.}
\label{gr_cr_w_0}
\end{figure}
\begin{figure}
\hspace{1.0cm}
\includegraphics[width=9.5cm,angle=0]{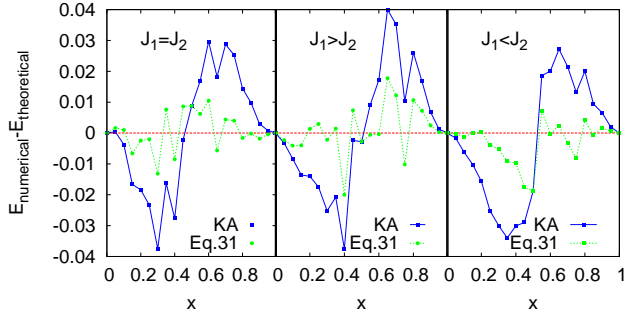}
\caption{Difference between the exit probability obtained from numerical simulation and
the one predicted by Kirkwood approximation (KA) and universal form (eq. \ref{alpha}) 
for FA model with $J_1=J_2$, $J_1>J_2$ and $J_1<J_2$.}
\label{fa}
\end{figure}

\subsection{ \bf $W_0$ model with next nearest neighbor interaction}
\label{sec:6}
In this case the flipping rate can be written as,
\begin{eqnarray*}
 w(\sigma_i)&=&\frac{1}{8}(3W_0+\frac{5}{2})-\frac{3}{16}\sigma_i(\sigma_{i+1}+\sigma_{i+2}+\sigma_{i-1}+\sigma_{i-2})\\
 &+&\frac{1}{16}\sigma_i(\sigma_{i+1}\sigma_{i+2}\sigma_{i-2}+\sigma_{i-1}\sigma_{i+2}\sigma_{i-2}\\
&+&\sigma_{i+1}\sigma_{i-1}\sigma_{i-2}+\sigma_{i+1}\sigma_{i+2}\sigma_{i-1})\\
&+&\frac{1}{8}(\frac{1}{2}-W_0)(\sigma_{i+1}\sigma_{i+2}+\sigma_{i+1}\sigma_{i-2}+\sigma_{i+1}\sigma_{i-1}\\
&+&\sigma_{i+2}\sigma_{i-2}+\sigma_{i+2}\sigma_{i-1}+\sigma_{i-1}\sigma_{i-2})\\
&+&\frac{3}{8}(W_0-\frac{1}{2})\sigma_{i+1}\sigma_{i+2}\sigma_{i-1}\sigma_{i-2}.
\end{eqnarray*}

\begin{figure}
\hspace{1.0cm}
\includegraphics[width=9.5cm,angle=0]{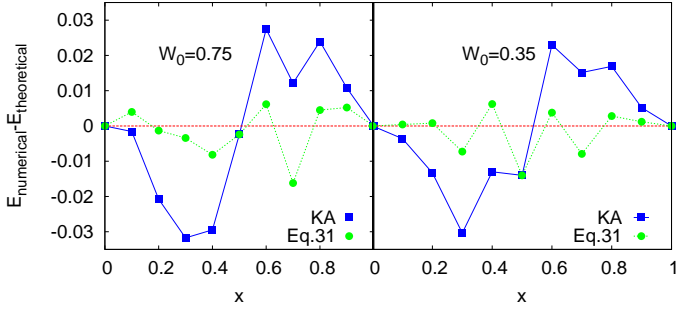}
\caption{Difference between the exit probability obtained from numerical simulation and
the one predicted by Kirkwood approximation (KA) and universal form (eq. \ref{alpha}) 
for $W_0$ model with next nearest neighbor interaction.}
\label{wmodel_r2}
\end{figure}

We have the rate equations for $m$ and $m_2$ using this flipping rate as,
\begin{equation}
\frac{dm}{dt}=m[(\frac{7}{8}-\frac{3}{4}W_0)+(\frac{3}{2}W_0-\frac{5}{4})m_2+\frac{3}{4}(\frac{1}{2}-W_0)m_2^2]
\label{eq1}
\end{equation}
and
\begin{equation}
 \frac{dm_2}{dt}=\frac{1}{4}[4-(1+m_2)^2].
 \label{eq2}
\end{equation}
Here we have factorized the $5$-point correlation function $\langle\sigma_i\sigma_{i+1}\sigma_{i+2}\sigma_{i-1}\sigma_{i-2}\rangle$ 
into $mm_2^2$ while writing down $\frac{dm}{dt}$.

Now solving the eq. (\ref{eq2}) we have $m_2=\frac{e^t-3C}{e^t+C}$,  
where $C=\frac{1-m_2(0)}{3+m_2(0)}$ and using this expression for $m_2$ we have from eq. (\ref{eq1}),
\begin{equation}
 m=C_1(1+Ce^{-t})^{(12W_0-8)}e^{\frac{C(6-12W_0)}{e^t+C}},
\end{equation}
where $C_1=\frac{m(0)}{(1+C)^{(12W_0-8)}\exp\left(\frac{C(6-12W_0)}{1+C}\right)}$.

Then following the same procedure as above we find $E(x)$ as,
\begin{eqnarray}
 E(x)&=&\frac{1}{2}[1+(2x-1)\nonumber\\
 &&(x^2-x+1)^{(12W_0-8)}
 e^{6x(1-x)(2W_0-1)}].
 \label{W_0_1}
\end{eqnarray}

\section{Simulation results and comparison}
\label{sec:7}
Earlier simulations \cite{psp} made for the five cases discussed in this paper suggested that the exit probability 
has a general form given by
\begin{equation}
E(x)=x^\alpha/[x^\alpha+(1-x)^\alpha] .
\label{alpha}
\end{equation}
In \cite{cast} only the $J_1=J_2$ case (eq. \ref{hamiltonian}) was considered which also yielded the same form 
with $\alpha=2$, agreeing with the result obtained later in \cite{psp}. We find that for the $W_0$ model with nearest neighbor interaction, 
both eq. (\ref{alpha}) and 
the Kirkwood approach eq. (\ref{W_0}) fit the data quite well (fig. \ref{wmodel}) and the differences  between these two 
forms and the simulation data plotted against 
$x$ (Fig. \ref{wmodel1}) 
do not show any systematic variation. Also the magnitude of the variations 
for both are of the same order ($\le 0.015$). 

Comparing the simulation results in the cases when interaction upto second neighbor is considered
with equations [\ref{exit}, \ref{exit1}, \ref{exit2}, \ref{W_0_1}] obtained using Kirkwood approximation, we 
note that there is a considerable difference between the two results and also the fact that the fitting with eq. (\ref{alpha}) 
is definitely better (Fig. \ref{gr_cr_w_0}). The difference between the simulation result and the analytical formula obtained using Kirkwood approximation 
plotted against $x$ shows systematic variations (Figures \ref{fa} and \ref{wmodel_r2}) as was observed in \cite{prado,new} for the 
nonlinear $q$ voter model. Here the magnitude of variations for Kirkwood approach ($\le 0.04$)
is much larger compared to those for the form given in eq. (\ref{alpha}) which have maximum value $\sim 0.02$. 
Although for the $J_1<J_2$ case we find a form for $E(x)$ (eq. \ref{exit2}) which has the same form as eq. (\ref{alpha}), 
the exponent $\alpha$ is quite different. For $J_1>J_2$ and $J_1<J_2$ 
the exponents $\alpha$ estimated were $1.85\pm0.03$ and $2.24\pm0.04$ respectively \cite{psp}. 

Simulation data shown in all the figures are from reference \cite{psp}. The simulation data do not show any appreciable 
system size dependence.

\section{Summary and discussions}
\label{sec:8}
In this paper, we have derived analytical expressions for the exit probability
for generalised Ising Glauber models in one dimension. We have used Kirkwood
approximation  together with an additional approximation following \cite{slan,kirk} to derive the results.
The models considered here involve either nearest neighbour interaction or both
nearest and next nearest neighbour interactions. The analytical results are compared with earlier results obtained by
numerical simulations reported in \cite{cast,psp}.

 The Kirkwood approximation scheme presented here appears to be very efficient for models with 
 nearest neighbour interaction. Although 
mathematically eq. (\ref{alpha}) and eq. (\ref{W_0}) cannot be reduced to the same form, it is interesting to note that
 their behaviour is almost indistinguishable for any $W_0$ in general 
 (figures \ref{wmodel} and \ref{wmodel1}) and it is difficult to conclude 
 which expression fits the data better. For $W_0>\frac{1}{2}$, one can simplify eq. (\ref{W_0}) as the 
exponential function is less than $1$. We take the example of $W_0=1$. Here the numerical simulation fits 
eq. (\ref{alpha}) with $\alpha=0.7$. Eq. (\ref{W_0}) for $W_0=1$ can be written as
\begin{eqnarray}
\label{comp}
 E(x)&=& \frac{1}{2}[1+(2x-1)e^{-2x(1-x)}]\nonumber\\
     &=& \frac{1}{2}[1+\frac{2x-1}{1+2x(1-x)+2x^2(1-x)^2}].
\end{eqnarray}
The right hand side of eq. (\ref{comp}) can be reduced to the form $$\frac{x^{0.7}}{x^{0.7}+(1-x)^{0.7}[(\frac{1-x}{x})^{0.3}
(\frac{1+x+x^2-x^3}{x^3-2x^2+2})]}.$$ The term within the third bracket is found to be very close to $1$ 
except for $x=0$ and $x=1$ such that we find that effectively one gets a form as in eq. (\ref{alpha}) with $\alpha=0.7$. 
However in general the two forms cannot be shown to be identical.

One may try to analyse the reasons behind the failure of KA (along with eq. (\ref{ka3})) for cases with 
second neighbor interaction. The first step, that of decomposing three 
spin correlations has been employed in both nearest neighbor and next nearest neighbor cases. 
In fact in the $W_0$ model with next nearest neighbor interaction, we have extended 
this for $5$ spin correlations as well.
The other assumption of approximating any $2$-point correlation as $m_2$ (eq. {\ref{ka3}) 
has also been employed in both; however $n\le2$ for the nearest neighbor case (eq. (\ref{near2})) while 
for the next nearest neighbor cases, we use this approximation for $n=3$ and $4$ as well (eq. (\ref{nnn})).
Using this approximation (eq. {\ref{ka3}) for values of $n>2$ as well as using equations (\ref{ka1}) and (\ref{ka2}) 
coming from KA may appear to be conflicting unless a mean field scenario is valid. The fact is the combination of approximations (\ref{ka1}), (\ref{ka2}) and (\ref{ka3}) 
yield results which deviate considerably in the next nearest neighbor cases. 
Apparently this is consistent with the fact that applying equation (\ref{ka3}) for $n>2$ 
is responsible for the error and the mean field scenario cannot be true as the models are 
short range in nature (even when the range of interaction is extended to next nearest neighbor). 
A remedy for this could be to introduce new variables $m_n=\langle\sigma_i\sigma_{i+n}\rangle$ for $n>2$, however, that would complicate the method 
to a large extent. At present, we believe this is the best possible way to study $E(x)$ analytically.

If we look at figures \ref{fa} and \ref{wmodel_r2}, we see that deviations $(|E_{numerical}-E_{theoretical}|)$ for the FA model 
are more compared to the $W_0$ model with next nearest neighbor interaction. The difference in these two models 
is that $W_0$ model is more deterministic in nature. Although a further decoupling scheme is employed for the 
$W_0$ model, the fact that the deviations are still less than that in FA indicates that this 
is not affecting the results to a large extent. The stochasticity in FA therefore 
appears to be responsible for the larger errors.

In conclusion, we find that although the KA is not very well understood, it is possible 
to arrive at analytical expressions for exit probability using it in various short range models. The 
method works very well for models with nearest neighbor interaction as shown by the agreement 
with numerical results. Also, we show that the question of existence of a universal expression for exit probability \cite{psp},
even for nearest neighbor models, is reopened.
For long ranged models, however, one needs to improve the approximation which may be a 
topic for future research.

\begin{acknowledgements}
We acknowledge Soumyajyoti Biswas for a critical reading of the draft version of the manuscript. We also thank 
Soham Biswas for discussions.
PR acknowledges financial support from  University Grant Commission. PS acknowledges financial support from CSIR project.

\end{acknowledgements}

\section*{Appendix}
\appendix
\numberwithin{equation}{section}

\section{Detailed derivation of spin flip rate for $W_0$ model with nearest neighbor interaction}
\label{Appendix A}
The most general spin flip rate for models with only nearest neighbor interaction 
satisfies the following natural requirements: \cite{krapivsky}
\begin{enumerate}
 \item Locality: Since it involves only nearest neighbor interaction, the spin flip rates should 
 also depend only on the nearest neighbor of each spin. Thus, $w(\sigma_i)=w(\sigma_{i-1},\sigma_i,\sigma_{i+1})$.
 \item Left/right symmetry: Invariance under the interchange $i-1\longrightarrow i+1$.
\end{enumerate}
The most general flip rate that satisfies these conditions has the form,
\begin{equation}
 w(\sigma_i)=A+B\sigma_i(\sigma_{i-1}+\sigma_{i+1})+C\sigma_{i-1}\sigma_{i+1}.
\end{equation}
Imposing the conditions of spin flipping for $W_0$ model with nearest neighbor interaction we have the 
following three equations:
\begin{equation}
 A-2B+C=1,
\end{equation}
\begin{equation}
 A+2B+C=0
\end{equation}
and
\begin{equation}
 A-C=W_0.
\end{equation}
Solving the above three equations 
the values of the constants are found to be $A=\frac{1}{4}(1+2W_0)$,  $B=-\frac{1}{4}$, $C=\frac{1}{4}(1-2W_0)$.
Thus the spin flip rate for $W_0$ model with nearest neighbor interaction has a form given in equation (\ref{flipw}).

\section{Detailed derivation of spin flip rate for model with next nearest neighbor interaction}
\label{Appendix B}
Invoking locality and left/right symmetry the most general spin flip rate for models with next nearest 
neighbor interaction has the form,
\begin{eqnarray}
 w(\sigma_i)&=& A+B\sigma_i(\sigma_{i+1}+\sigma_{i-1})+C\sigma_i(\sigma_{i+2}+\sigma_{i-2})\nonumber\\
 &+& D\sigma_i(\sigma_{i+1}\sigma_{i+2}\sigma_{i-2}+\sigma_{i+1}\sigma_{i+2}\sigma_{i-1}+\sigma_{i+2}\sigma_{i-1}\sigma_{i-2}\nonumber\\
 &+& \sigma_{i+1}\sigma_{i-1}\sigma_{i-2})+E(\sigma_{i+1}\sigma_{i+2}+\sigma_{i+1}\sigma_{i-2}+\sigma_{i+1}\sigma_{i-1}\nonumber\\
 &+& \sigma_{i+2}\sigma_{i-2}+\sigma_{i+2}\sigma_{i-1}+\sigma_{i-2}\sigma_{i-1})+F\sigma_{i+1}\sigma_{i+2}\sigma_{i-1}\sigma_{i-2}.
\end{eqnarray}

Using the conditions of spin flipping for FA model with $J_1=J_2$, we have the following equations,
\begin{equation}
A+2B+2C+4D+6E+F=0, 
\end{equation}
\begin{equation}
 A-2B-2C-4D+6E+F=0, 
\end{equation}
\begin{equation}
A+2B-2C-2E+F=\frac{1}{2}, 
\end{equation}
\begin{equation}
 A-2B+2C-2E+F=\frac{1}{2},
\end{equation}
\begin{equation}
A-2B+2D-F=1
\end{equation}
and
\begin{equation}
 A+2B-2D-F=0.
\end{equation}
Solving the above six equations we have, 
$A=\frac{1}{2}$, $B=C=-\frac{3}{16}$, $D=\frac{1}{16}$, $E=F=0$,
which leads to the equation (\ref{flipfa}) for the spin flip rate of FA model with $J_1=J_2$.
Similarly the spin flip rates for other models with next nearest neighbor interaction can also be obtained.


\end{document}